\documentclass[iop]{emulateapj}
\usepackage{icomma, enumitem, graphicx, threeparttable}
\newcommand{\hi}{H\,{\scriptsize I}}
\newcommand{\hii}{H\,{\scriptsize II}}
\newcommand{\heii}{He\,{\scriptsize II}}
\newcommand{\nii}{[N\,{\scriptsize II}]}
\newcommand{\ovi}{O\,{\scriptsize VI}}
\newcommand{\ha}{H$\alpha$}

\newcommand{\kpc}{\,{\rm kpc}}
\newcommand{\kms}{\,km\,s$^{-1}$}
\newcommand{\myr}{\,$\rm M_{\odot}\,{\rm yr}^{-1}$}
\newcommand{\ro}{\,$\rm R_{\odot}$}
\newcommand{\mo}{\,$\rm M_{\odot}$}
\newcommand{\lo}{\,$\rm L_{\odot}$}

\newcommand{\cmtt}{\,cm$^{3}$}
\newcommand{\cmd}{\,cm$^{-2}$}

\newcommand{\ecsa}{$\rm\,erg\,cm^{-2}\,s^{-1}\,\AA^{-1}$}

\slugcomment{Draft version June 22, 2015}

\shorttitle{Mass-loss rate from Raman scattering}
\shortauthors{M. Seker\'a\v{s} and A. Skopal}

\begin{document}

\title{Mass-loss rate by the Mira in the symbiotic binary 
       V1016~Cygni
       from Raman scattering}

\author{M. Seker\'a\v{s} and A. Skopal}
\affil{Astronomical Institute, Slovak Academy of Sciences,
059\,60 Tatransk\'{a} Lomnica, Slovakia; sekeras@ta3.sk (MS), skopal@ta3.sk (AS)}

\begin{abstract}
The mass-loss rate from Mira variables represents a key parameter 
in our understanding of their evolutionary tracks. We introduce 
a method for determining the mass-loss rate from the Mira 
component in D-type symbiotic binaries via the Raman scattering 
on atomic hydrogen in the wind from the giant. 
Using our method, we investigated Raman 
\heii\,$\lambda1025\rightarrow\lambda6545$ conversion 
in the spectrum of the symbiotic Mira V1016~Cyg. 
We determined its efficiency, 
$\eta = $0.102, 0.148, and the corresponding mass-loss 
rate, $\dot{M} = 2.0^{+0.1}_{-0.2}\times10^{-6}$, 
$2.7^{+0.2}_{-0.1}\times10^{-6}$\myr, using our spectra 
from 2006 April and 2007 July, respectively. 
Our values of $\dot{M}$ that we derived from
Raman scattering are comparable with those obtained 
independently by other methods. 
Applying the method to other Mira--white dwarf binary systems 
can provide a necessary constraint in the calculation 
of asymptotic giant branch (AGB) evolution. 
\end{abstract}


\section{Introduction}
\label{sect:intro}

Symbiotic stars are the widest interacting binaries that comprise 
a late-type cool component, which is a red/yellow giant or a Mira 
variable and a hot compact star, in most cases a white dwarf (WD). 
The WD accretes a fraction of the stellar wind from the giant, 
which makes it very hot ($\approx 10^{5}\,\rm K$) and luminous 
($\approx 10^{2}-10^{4}\, \rm L_{\odot}$), and thus capable of 
ionizing the neutral wind from the giant, giving rise to the 
so-called symbiotic nebula \citep[][ hereafter STB]{stb84}. 
Assuming a stationary binary during the quiescent phase, STB 
found that the neutral region usually has the shape of 
a cone, with the giant below its top, facing the WD. 
According to the spectral energy distribution in the IR region, 
we distinguish S-type (stellar) and D-type (dusty) symbiotic stars 
\citep[][]{wa75}. The former is represented by a stellar type of the 
IR continuum from a normal giant, whereas the latter contains an 
additional emission from the dust produced by an evolved Mira-type 
variable. D-type symbiotics are therefore called symbiotic Miras.

The environment of symbiotic stars is very suitable for observing the
effects of Thomson, Rayleigh, and Raman-scattering processes 
\citep[e.g.][]{schmid97}. Thomson scattering by free electrons 
arises in the symbiotic nebula. Its effect on the broadening of 
strong emission lines was recently investigated by \cite{ss12}. 
The effects of Raman and Rayleigh scattering processes are observable 
when the radiation from/around the WD passes through the neutral 
part of the giant wind. In these processes a photon excites an atom 
from its ground state to an intermediate state, which is immediately 
stabilized by a transition to a true bound state. If this is 
followed by the immediate reemission of a photon of the same 
wavelength, we talk about Rayleigh scattering. 
If the stabilizing transition results in emitting a photon of 
a different frequency, we talk about Raman scattering. Raman 
scattering as a diagnostic tool in astrophysics was 
outlined by \cite{nsv89}. 

The most famous example of Raman-scattered lines in the 
spectrum of symbiotic stars is represented by broad emission 
features at 6825 and 7082\,\AA\ that are formed when 
the photons of \ovi\ 1032 and 1038\,\AA\ emission lines 
are scattered by neutral hydrogen atoms in the wind from 
the giant \citep[]{sch89,sea99}. 
%
%
\begin{figure*}
\centering
\resizebox{0.95\textwidth}{!}
          {\includegraphics[angle=-90]{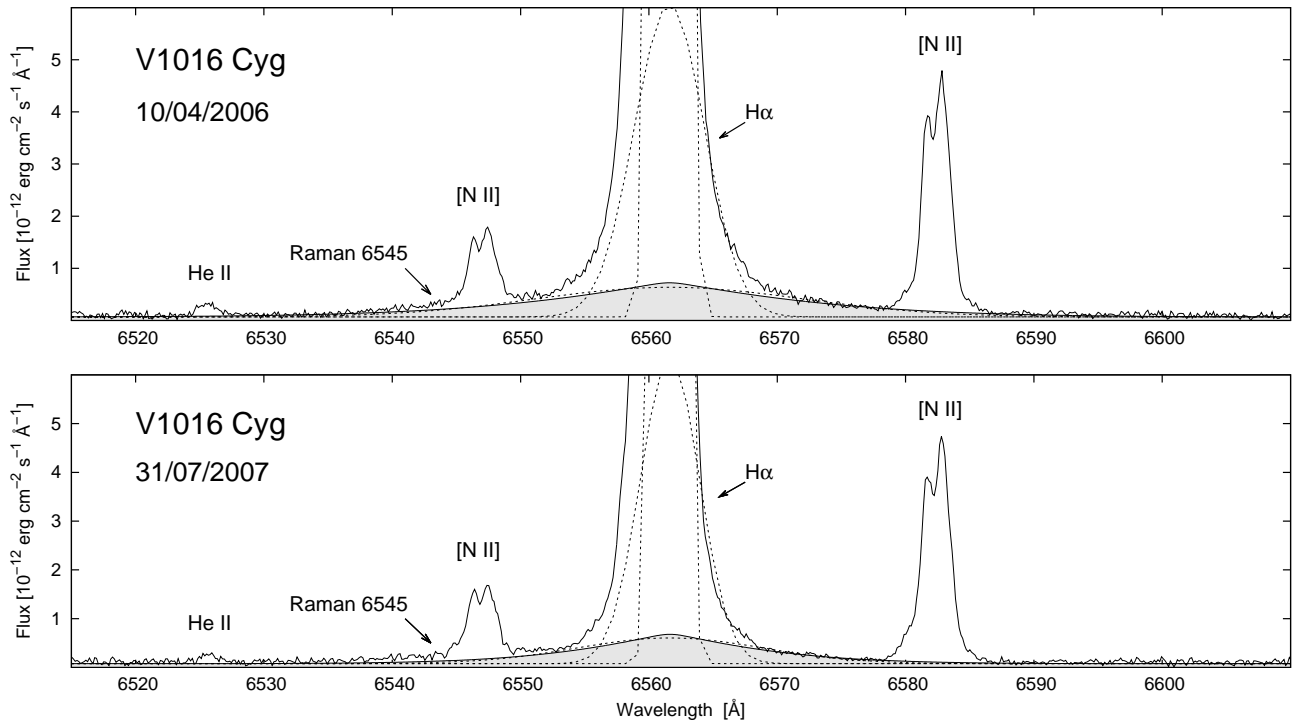}}
\caption[]{
Our spectra of V1016~Cyg covering the vicinity of the \ha\ line. 
Its profile was approximated by three Gaussians (dotted lines). 
The broadest Gaussian was matched by a scattering of \ha\ photons 
on free electrons in the symbiotic nebula (shaded area). 
A slightly enhanced blue \ha\ wing around the emission line 
\nii\,6548\,\AA\ indicates the possible presence of Raman 
6545\,\AA\ emission. 
}
\label{fig:spec}
\end{figure*}
Also, transitions between the levels of even principal quantum numbers 
in the \heii\ atom produce photons that can be easily scattered 
onto neutral hydrogen atoms because their wavelengths are near 
to the \hi\ Lyman transitions, where the scattering cross-section 
is relatively large \citep[e.g.][]{lele97,lee12}. 
In particular, the Raman emission at 6545\,\AA, produced by 
\heii\,1025\,\AA\ scattering, was observed in some planetary 
nebulae and symbiotic stars 
\citep[e.g.][]{pea97,lkb01,lee03,lee06}. Its cross-section is 
$8.05\times10^{-22}\,\rm cm^{2}$ \citep{lee09}, which means 
that it operates in scattering regions with 
\hi\ column density, $N_{\rm HI} \ga 10^{21}$\,\cmd. 
In symbiotic binaries the neutral material is supplied 
into the binary environment by the cool giant. Assuming the STB 
geometry of the neutral region, the Raman conversion process 
is related to its size, which depends on the mass-loss rate from 
the giant, $\dot M$ (see STB). Therefore, Raman-scattered lines 
can provide an independent determination of $\dot M$. 

In this paper we investigate the \heii\ Raman-scattered emission 
at 6545\,\AA\ in our spectra of D-type symbiotic star V1016~Cyg. 
In 1964, V1016~Cyg underwent a nova-like outburst \citep{mc65}. 
Its brightness is very slowly fading from its peak 
\textit{V}$\sim$10.8\,mag in 1967-1970 to \textit{V}$\sim$11.9\,mag in 2007 
\citep[e.g.][]{mn94,pea02,aea08}. The binary is very extended, 
as given by its orbital period, estimated to be 6 -- 544 years 
\citep[e.g.][ and the references therein]{pea02}. Analysis of 
the ultraviolet spectrum revealed the presence of a very hot 
($\approx$150,000\,K) and luminous ($\approx$30,000\lo) 
WD, whose radiation gives rise to a strong and extended symbiotic 
nebula by ionizing a fraction of the neutral wind from its 
Mira-type companion \citep[e.g.][]{ns81,ss90,mn94}. 

In Section~\ref{sect:dis} we describe a method for isolating
Raman-scattered emission at 6545\,\AA\ from the spectrum, and in 
Section~\ref{sect:eff} we determine the efficiency of the 
corresponding \heii\,$\lambda1025\rightarrow\lambda6545$ 
conversion. Assuming the basic ionization structure of 
V1016~Cyg (Section~3.3.1) we estimate $\dot M$ from its cool 
component (Section~3.3.2). A discussion and our conclusions
are found in Sections~\ref{sect:disc} and \ref{sect:con}, respectively. 

\section{Observations}
\label{sect:obs}
Four optical spectra (6420--6710\,\AA) of the symbiotic star 
V1016~Cyg were obtained at the David Dunlap Observatory of the
University of Toronto by an 1.88-m telescope equipped with 
a single-dispersion slit spectrograph and the Jobin Yovon 
Horiba CCD detector (2048$\times$512 pixels) at the Cassegrain 
focus. Two of them were obtained on 2006 April 10, with exposure 
times of 120 and 60\,s, and two were obtained on 2007 July 31,
with 100 and 60\,s exposures. 
We combined the two spectra from each date to find
both the continuum and the strong \ha\ line in one spectrum. 
The resulting spectra were processed with basic procedures using the
IRAF software package. 

Near-simultaneous CCD photometry was obtained at the Star\'{a} 
Lesn\'{a} observatory using the {\sc SBIG ST10 MXE} CCD camera 
with the chip 2184$\times$1472 pixels and the $UBV(RI)_{\rm C}$ 
Johnson-Cousins filter set mounted at the Newtonian focus of 
a 0.5\,m telescope \citep[see][ in detail]{pv05}. 
Magnitudes $U$ = 11.10, $B$ = 11.96, $V$ = 11.92, 
$R_{\rm C}$ = 10.55, $I_{\rm C}$ = 11.03, and 
$U$ = 11.21, $B$ = 12.09, $V$ = 12.14, $R_{\rm C}$ = 10.73, 
$I_{\rm C}$ = 11.15, were measured on 2006 April 8 and 2007
July 7, respectively. 

Arbitrary units of the spectra were converted to fluxes in 
\ecsa\ using $V$ and $R_{\rm C}$ magnitudes that were corrected for 
emission lines \citep[see][]{sk07}. Their contribution 
to the true continuum was $\Delta R_{\rm C}$ = -1.51 and 
-1.26\,mag in our spectra. 
The correction in the $V$ filter was adopted from Table~2 of 
\cite{sk07}. Magnitudes were converted to fluxes according 
to the calibration of \cite{hk82}. 
Finally, the spectra were corrected for interstellar reddening 
according to \cite{ccm89} using the color excess 
$E_{\rm B-V}$ = 0.28\,mag \citep{ns81}.

\section{Analysis and results}
\label{sect:a&r}
\subsection{Disentangling of Emission Lines}
\label{sect:dis}

\begin{figure}
\centering
\resizebox{0.45\textwidth}{!}
          {\includegraphics[angle=-90]{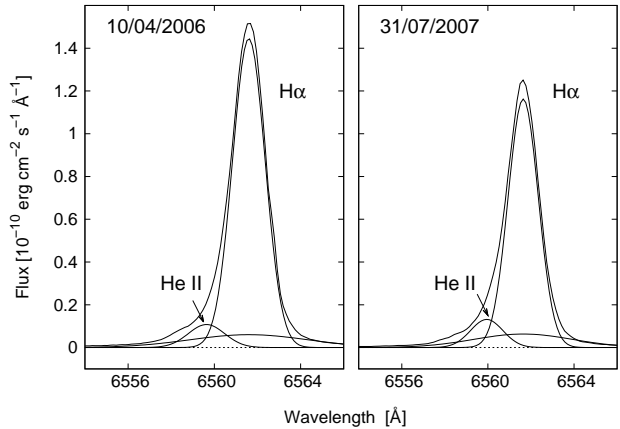}}
\caption[]{
Observed profile of the \ha\ line (thick line). 
Its components are denoted by thin lines, including 
the \heii\,6560\,\AA\ line. 
           }
\label{fig:ha}
\end{figure}

Both spectra were dominated by a strong \ha\ line with broad 
wings, with a total equivalent width of 4064 and 
2764\,\AA, and FWHMs of 1.8 and 1.6\,\AA\ in 2006 April and
2007 July (see Figs.~\ref{fig:spec} and \ref{fig:ha}), respectively. 

\begin{figure}
\centering
\vspace*{10pt}
\resizebox{0.45\textwidth}{!}
          {\includegraphics[angle=-90]{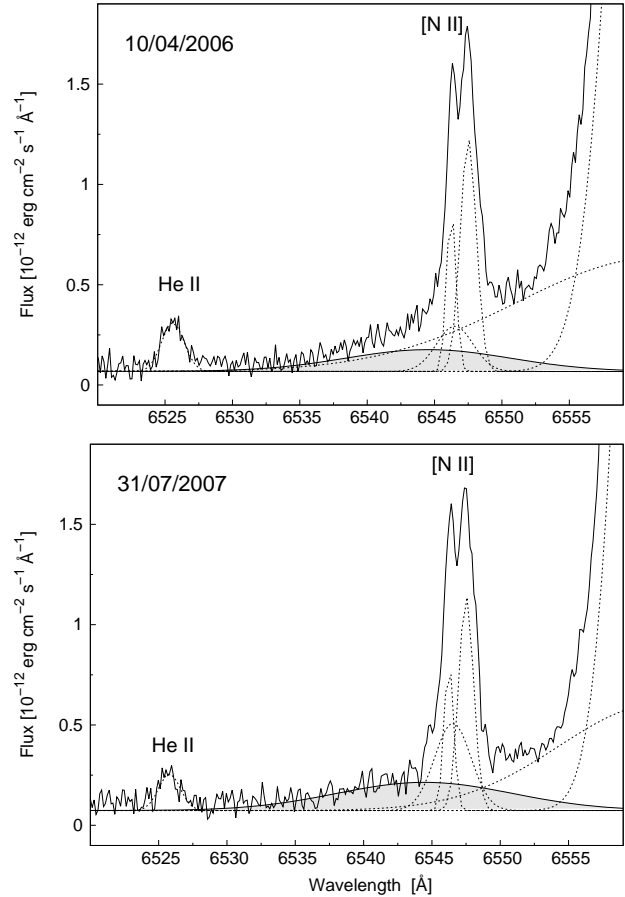}}
\caption[]{
Raman 6545\,\AA\ emission (shaded area) blended 
in the wings of \ha\ and \nii\ 6548\,\AA\ lines. 
Their profiles were approximated by Gaussian functions 
(dotted lines, Section~3.1). 
                 }
\label{fig:ram}
\end{figure}

\begin{figure}
\centering
\resizebox{0.45\textwidth}{!}
          {\includegraphics[angle=-90]{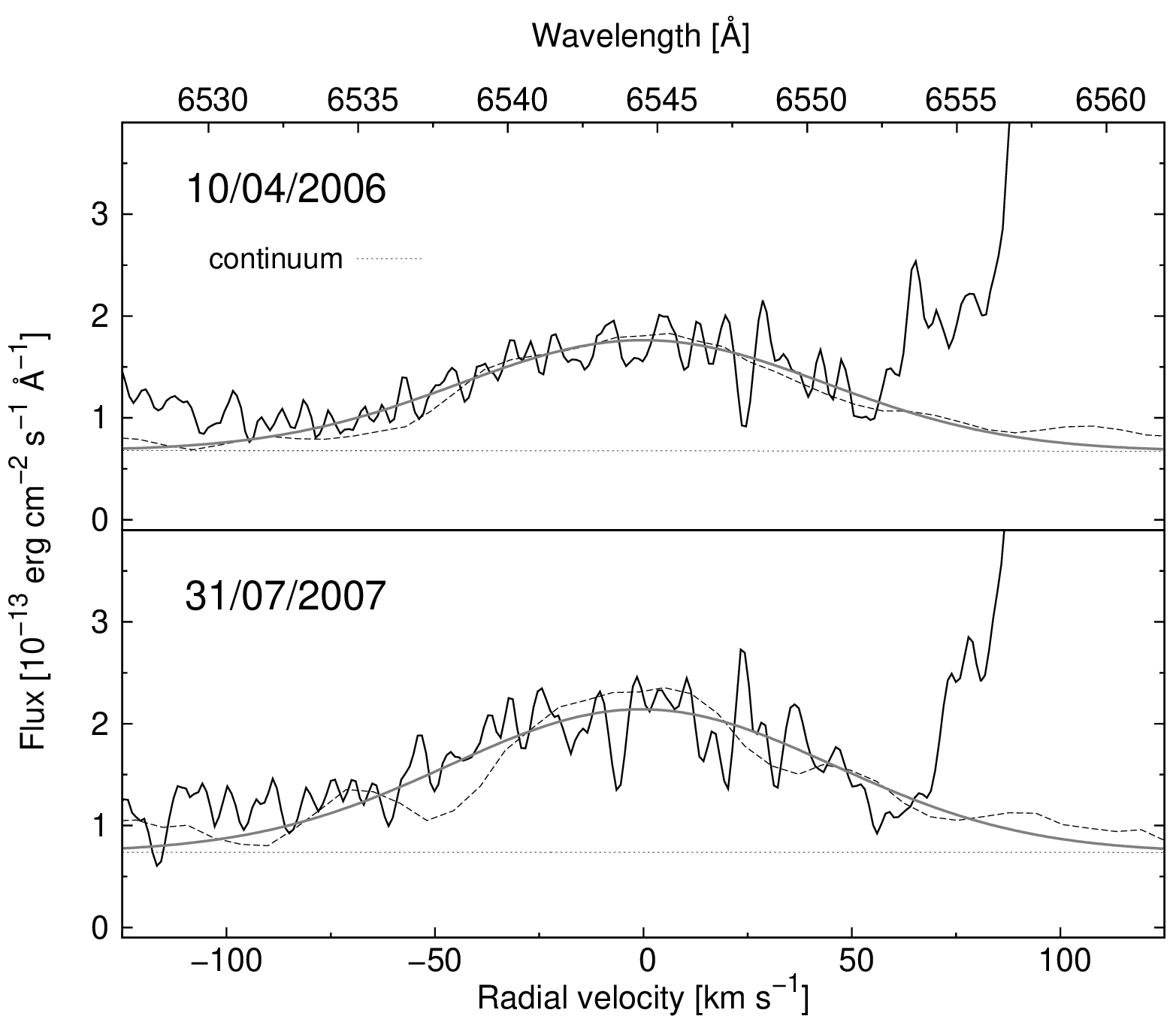}}
\caption[]{
Isolated Raman-scattered 6545\,\AA\ emission feature 
(thick line) matched by the Gaussian function (gray line). 
It is compared with the \heii\,6527\,\AA\ line profile (dashed line). 
The radial velocity scale corresponds to the velocity space 
of the ordinary lines. The flux of the lines is normalized 
to the same total flux (see Section~4.3).
           }
\label{fig:isol}
\end{figure}

The slightly enhanced blue wing of \ha\ around \nii\,6548\,\AA\ suggests 
the presence of Raman 6545\,\AA\ emission. To isolate it, we 
proceeded similarly as \cite{lee03}. First, we approximated the 
continuum by a linear function. Then we removed contributions 
from the \ha\ wing, and the \heii\,6560\,\AA\ and \nii\,6548\,\AA\ lines. 
All profiles were approximated by Gaussian functions 
(Figs.~\ref{fig:spec}--\ref{fig:ram}). 
\begin{itemize}[noitemsep]
\item [i.] The \ha\ line was fitted by three Gaussians with the same 
central wavelength. The first curve fits the strong and relatively 
narrow core of the line ($FWHM \sim 1.7$\,\AA), whose origin can 
be connected with the ionized wind from the giant. The second curve 
extracts a fainter but broader component ($FWHM \sim 6$\,\AA), 
probably formed in the wind from the burning WD \citep[see][]{sk06}. 
The third Gaussian matches the faint and very extended wings 
($FWHM \sim 21$\,\AA), which can be caused by a scattering 
of free electrons \citep[see][]{ss12}, but a contribution 
from the Raman scattering of the continuum around Ly$\beta$ 
cannot be excluded \citep[e.g.][]{lee00}. 
\item [ii.] The \nii\,6583 and 6548\,\AA\ lines consist of three 
components; a double-peaked core with a broad base. 
Therefore, we fitted their profiles with three Gaussians
and subtracted the \nii\,6548\,\AA\ line using the ratio 
$F_{6583}/F_{6548} \sim 3$ \citep[e.g.][]{sz00}. 
\item [iii.] The \heii\,6560\,\AA\ line was fitted by a single 
Gaussian to tune the composite \ha\ line profile at its blue 
side (Figure~\ref{fig:ha}). 
\end{itemize}
Finally, subtracting the above mentioned contributions from 
the spectrum, we isolated the Raman 6545\,\AA\ emission 
(see Figs.~\ref{fig:ram} and \ref{fig:isol}). Parameters of all 
fitted emission lines are in Table ~\ref{tab:01}. 
%
%
%
\\
%
%
%
%
%
\begin{table*}
\begin{center}
\begin{threeparttable}
\caption{
Gaussian fit parameters of the Used Emission Lines (see Section~3.1). 
         }							
\begin{tabular}{lcccccc}
\tableline\tableline
\\
             & \multicolumn{3}{c}{2006 Apr}  & \multicolumn{3}{c}{2007 Jul} \\
\hline
Emission Line&$\lambda_{0}$ &${F_{0}}$ &FWHM &$\lambda_{0}$ &${F_{0}}$&FWHM  \\
             &(\AA)         &(\ecsa)   &(\AA)&(\AA)         &(\ecsa)  &(\AA) \\
\hline
\heii
     &6525.6  &2.4$\times10^{-13}$  &1.9  &6525.7& 1.8$\times10^{-13}$ &1.9  \\
Raman \heii\,6545  
     &6544.5  &1.1$\times10^{-13}$  &14.5 &6544.5& 1.4$\times10^{-13}$ &15.2 \\
\nii &6546.2  &8.2$\times10^{-13}$  &0.8  &6546.2& 7.5$\times10^{-13}$ &~0.8 \\
     &6546.5  &2.3$\times10^{-13}$  &3.2  &6546.5& 4.4$\times10^{-13}$ &3.1  \\
     &6547.5  &1.2$\times10^{-12}$  &1.4  &6547.5& 1.1$\times10^{-12}$ &1.3  \\
\heii&6559.6  &1.1$\times10^{-11}$  &2.0  &6559.9& 1.1$\times10^{-11}$ &1.8  \\
\ha  &6561.6  &1.4$\times10^{-10}$  &1.8  &6561.6& 1.2$\times10^{-10}$ &1.6  \\
     &6561.6  &5.7$\times10^{-13}$  &24.7 &6561.6& 5.3$\times10^{-13}$ &17.3 \\
     &6561.6  &6.0$\times10^{-12}$  &6.4  &6561.6& 6.3$\times10^{-12}$ &5.4  \\
\nii &6581.6  &2.4$\times10^{-12}$  &0.8  &6581.6& 2.2$\times10^{-12}$ &0.8  \\
     &6582.3  &1.0$\times10^{-12}$  &3.7  &6582.1& 1.5$\times10^{-12}$ &3.6  \\																									
     &6582.9  &3.5$\times10^{-12}$  &1.4  &6582.9& 3.2$\times10^{-12}$ &1.3  \\
\tableline
\label{tab:01}
\end{tabular}
\begin{tablenotes}
\item {\bf Note.} Denotation of parameters is given by the Gaussian function, 
$F_{0}\exp(-0.5(\lambda-\lambda_{0})^2/\sigma^2)$ 
and FWHM = 2$\sigma\sqrt{2\ln 2}$. 
\end{tablenotes}
\end{threeparttable}
\end{center}
\end{table*}
%
%

\subsection{Raman-scattered \heii\,6545\,\AA\ Emission}
\label{sect:ram}

According to the energy conservation of the
Raman transition $\lambda 1025\rightarrow \lambda 6545$, 
\begin {equation} 
   \lambda^{-1}_{\rm R}=\lambda^{-1}_{i}-\lambda^{-1}_{\rm Ly\alpha},
\label{eq:energy}
\end {equation}
where $\lambda_{\rm R}$ is the wavelength of the Raman-scattered 
emission at 6545\,\AA, $\lambda_{\rm i}$ is the wavelength of the 
incident photons of the \heii\,1025\,\AA\ line, and 
$\lambda_{\rm Ly\alpha}$ is the wavelength of the Ly$\alpha$ 
line. Differentiating Equation~(\ref{eq:energy}), we get, 
\begin{equation} 
  \frac{\Delta\lambda_{\rm R}}{\lambda_{\rm R}} = 
              \frac{\lambda_{R}}{\lambda_{i}}
              \frac{\Delta\lambda_{\rm i}}{\lambda_{i}}.
\label{eq:dl}
\end{equation}
Hence the Raman-scattered feature is broadened with respect to 
the incident emission line by a factor of 
$(\lambda_{\rm R}/\lambda_{\rm i})^{2} \approx 40.8$ in the
wavelengths space or $\lambda_{\rm R}/\lambda_{\rm i}\approx 6.4$ 
in the radial velocity space. 

\cite{lee06} calculated the central wavelengths of the 
\heii\,6560\,\AA\ and Raman 6545\,\AA\ emission lines as 
$\lambda_{\rm He II} = 6560.13$\,\AA\ and 
$\lambda_{\rm Ram} = 6544.53$\,\AA, setting the position of 
the Raman 6545\,\AA\ emission to 15.6\,\AA\ blueward of the 
\heii\,6560\,\AA\ line. However, the exact position depends 
on several factors, e.g. a relative motion of the scattering 
and emitting regions and/or the column density of the neutral 
hydrogen \citep{jl04}. The emission feature, which we extracted from 
the blue wing of the \ha\ line, is shifted by 15.1 and 15.4\,\AA\ 
with respect to the \heii\,6560\,\AA\ line in our spectra 
from 2006 and 2007, respectively. 
Therefore, its position and the width (Table~\ref{tab:01}, 
Figure~\ref{fig:isol}) imply that it is indeed the Raman-scattered 
\heii\,1025\,\AA\ line at 6545\,\AA. 

\subsubsection{Efficiency of Raman \heii\,$\lambda1025 
               \rightarrow \lambda6545$ Scattering} 
\label{sect:eff}

The Raman-scattering efficiency, $\eta$, is defined as the photon 
ratio between the Raman-scattered, $N_{6545}$, and the original 
\heii\,1025\,\AA\ line photons, $N_{1025}$, 
\begin {equation} 
  \eta = \frac{N_{6545}}{N_{1025}}.
\label{eq:eta1}
\end {equation}
According to \cite{lee03}, this can be expressed as 
\begin {equation} 
\eta=\frac{F_{6545}/h\nu_{6545}}{F_{1025}/h\nu_{1025}}
 =\frac{F_{6545}/h\nu_{6545}}{F_{6560}/h\nu_{6560}} \quad
	 \frac{F_{6560}/h\nu_{6560}}{F_{1025}/h\nu_{1025}},
\label{eq:eta2}
\end {equation}
where $F_{6560}$ is the flux of the \heii\,6560\,\AA\ 
emission line. This approach takes advantage of the fact
that even though we have no observation of the original 
\heii\,1025\,\AA\ line, we can still determine the efficiency 
using the observed \heii\,6560\,\AA\ line located close to 
Raman 6545\,\AA\ emission (Figure~\ref{fig:ha}). According to 
\cite{g97}, under case B of recombination for the electron 
temperature $T_{\rm e} = 20,000$\,K, 
theoretical fluxes of the referred \heii\ lines in 
Equation~(\ref{eq:eta2}) are $F_{6560} = 0.135\,F_{4686}$ and 
$F_{1025} = 0.618\,F_{4686}$, where $F_{4686}$ is the total flux 
of the \heii\,4686\,\AA\ line. 
Assuming an isotropic \heii-emitting region, the Raman-scattering
efficiency $\eta = 0.102$ and 0.148 in 2006 April
and 2007 July, respectively. For $T_{\rm e} = 10,000$\,K, $\eta$ will 
increase to 0.129 and 0.187, respectively. 

\subsubsection{Covering factor of the scattering region}
\label{sect:cs}

\cite{lee09} determined the cross-section of the Raman \heii\, 
$\lambda 1025 \rightarrow \lambda 6545$ conversion to
$\sigma_{\rm Ram} = 8.05\times10^{-22}\rm cm^{2}$. 
The corresponding optical depth is 
\begin{equation}
  \tau_{\rm Ram} = N_{\rm HI}\,\sigma_{\rm Ram},
\label{eq:tau}
\end{equation}
where $N_{\rm HI}$ is the column density of the neutral hydrogen 
region along the line of the incident \heii\,1025\,\AA\ photons. 
This implies that for $N_{\rm HI} > 1.24\times 10^{21}$\cmd\ 
(i.e. $\tau_{\rm Ram} > 1$), the \hi\ region becomes optically 
thick for Raman scattering. Performing a detailed Monte Carlo 
calculations, \cite{lee00} found that in the optically thick 
region the scattering efficiency reaches a maximum value of 0.6. 
This means that 60\% of all photons entering the optically 
thick scattering region, $N^{\rm e}_{1025}$, are converted to 
Raman 6545\,\AA\ photons, i.e., 
\begin{equation}
  0.6 \; N^{\rm e}_{1025} = N_{6545}.
\label{eq:06}
\end{equation}
A part of the \hi\ region, which is optically thick for the 
\heii\,1025\,\AA\ Raman-scattering, defines a covering 
factor $C_{\rm S}$ as (Equations~(\ref{eq:eta1}) and (\ref{eq:06})), 
\begin{equation}
  C_{\rm S} = \frac{N^{\rm e}_{1025}}{N_{1025}} = 
              \frac{\eta}{0.6},
\label{eq:cs}
\end{equation}
which represents a fraction of the sky, seen from the \heii\ 
emission zone, covered by the Raman scattering region. For 
the whole sphere, $C_{\rm S} = 1$. From our spectra, we 
determined the covering factors to be 0.171 and 0.248. 

\subsection{Mass-loss Rate from the Giant}
\label{sect:mass}
\subsubsection{Ionization Model}
\label{sect:im}

In the STB model, the boundary between the neutral and ionized 
part of the wind from the giant is represented by a locus of 
points, where the flux of hydrogen-ionizing photons from the 
hot star, $L_{\rm H}$, is balanced by the rate of 
recombinations in the giant's wind. In the WD-centered polar 
coordinate system ($s$, $\theta$), this equilibrium condition 
can be expressed as \citep[][]{nuv87}, 
\begin{equation}
  L_{\rm H} = 4\pi \int^{s_{\theta}}_0 (1+a({\rm He}))
              n^{2}_{\rm H}(s)\alpha_{\rm B}({\rm H},T_{\rm e})
              s^{2}{\rm d}s, 
\label{eq:lh}
\end{equation}
where $s_{\theta}$ is the distance to the \hi/\hii\ 
boundary from the WD for a given $\theta$, 
$a({\rm He})$ is the abundance by number of helium atoms, 
$n_{\rm H}(s)$ is the concentration of hydrogen atoms,
and $\alpha_{\rm B}({\rm H},T_{\rm e})$ is the total 
recombination coefficient for hydrogen atoms in case B 
(cm$^{3}$\,s$^{-1}$). 
The STB model assumes a stationary symbiotic binary with a 
spherically symmetric wind flowing from the giant at a constant 
velocity, v$_{\infty}$. Such a wind obeys the continuity 
equation
\begin{equation}
  \dot{M} = 4\pi \mu m_{\rm H}n_{\rm H}(r)v_{\infty},
\label{eq:cont}
\end{equation}
where $\dot M$ is the mass-loss rate from the giant, $r$ is the 
distance from its center, $\mu$ is the mean molecular weight, 
and $m_{\rm H}$ is the mass of the hydrogen atom. 
In the reference system of the WD, 
\begin{equation}
   r = \sqrt{s^{2} + p^{2}-2sp\cos\theta}, 
\label{eq:rs}
\end{equation}
where $p$ is the binary separation. Following these equations, 
the \hi/\hii\ interface is given by the solution of a parametric 
equation \citep[see][]{nuv87}, 
\begin{equation}
   f(u,\theta) = X_{\rm H^{+}},
\label{eq:f=X}
\end{equation}
where $u = s/p$, the parameter 
\begin{equation}
   X_{\rm H^{+}} = \frac{4\pi \mu^{2}m^{2}_{\rm H}}
                   {\alpha_{\rm B}(H,T_{e})(1+a(He))}p L_{\rm H}
                   \left(\frac{v_{\infty}}{\dot{M}}\right)^{2}, 
\label{eq:X}
\end{equation}
and the function 
\begin{equation}
   f(u,\theta) = \int^{u_{\theta}}_{0}\frac{u^{2}}
                  {(u^{2}-2u\cos \theta +1)^{2}}
                 \left(\frac{v_{\infty}}{v(r)}\right)^2 {\rm d}u, 
\label{eq:futh}
\end{equation}
where, instead of the constant velocity, we used the $\beta$-law 
wind velocity profile \citep[][]{lamcass99}
\begin{equation} 
   v(r) = v_{0}+(v_{\infty}-v_{0})
          \left(1-\frac{R_{g}}{r}\right)^{\beta}. 
\label{eq:vr}
\end{equation}
The initial velocity $v_{0} = v(R_{\rm g})$, where $R_{\rm g}$ 
is the radius of the giant, $v_{\infty}$ is the wind terminal 
velocity, and the parameter $\beta$ determines the steepness of 
the velocity law. The value of $v_{0}$ is often linked to 
the isothermal speed of sound \citep[][]{puls+08}. 

The setting of the STB model results in a symmetric ionization 
boundary with respect to the binary axis, whose shape is 
determined solely by the parameter $X_{\rm H^{+}}$. 
%
%
%
\begin{figure}
\begin{center}
\resizebox{0.45\textwidth}{!}
          {\includegraphics[angle=-90]{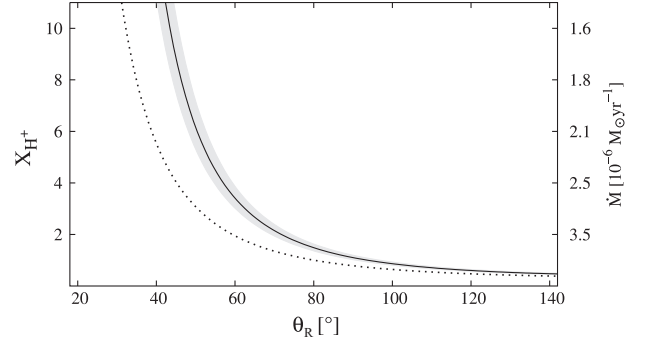}}
\end{center}
\caption[]{
Parameter $X_{\rm H^{+}}$ as a function of the angle 
$\theta_{\rm R}$, which limits the Raman-scattering \hi\ 
region (Equation~(\ref{eq:thr})). 
The dotted line represents the model for a constant wind 
velocity ($v_{\infty} = 11$\kms), while the solid line represents 
the $\beta$-law wind (Equation~(\ref{eq:vr})). 
The shaded area mirrors the uncertainty in $R_{\rm g}$ 
($\pm 73$\ro). The angle $\theta_{\rm R}$ determines 
$X_{\rm H^{+}}$ and $\dot M$ (Equation~(\ref{eq:xdotm})), 
the scale of which is given on the y2-axis. 
           }
\label{fig:xthr}
\end{figure}

\subsubsection{Mass-loss Rate from Raman Scattering}
\label{sect:x}

\begin{figure}
\begin{center}
\resizebox{0.4\textwidth}{!}
          {\includegraphics[angle=-90]{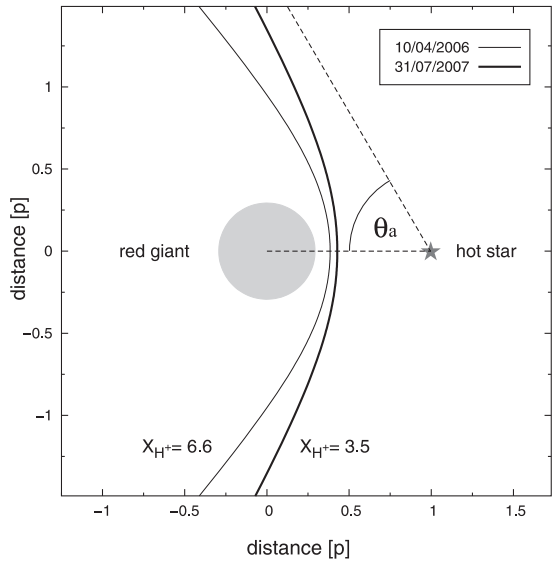}}
\end{center}
\caption[]{
Ionization boundaries determined from the Raman-scattered 
6545\,\AA\ emission feature observed in our spectra of 
V1016~Cyg. The neutral region is on the side of the boundary 
containing the giant. The asymptotic angle $\theta_{\rm a}$ 
(Equation~(\ref{eq:tha})) to the boundary $X_{\rm H^{+}} = 3.5$ 
is shown for comparison. 
           }
\label{fig:tha}
\end{figure}

According to the meaning of the covering factor 
(see Section~\ref{sect:cs}), we can express it via a solid angle 
$\Omega$, under which the initial \heii\ line photons, located 
mostly in the vicinity of the WD, can see the scattering 
region, i.e. 
\begin{equation}
  C_{\rm S} = \frac{\Omega}{4\pi} = 
              \frac{1-\cos\theta_{\rm R}}{2}. 
\label{eq:omega}              
\end{equation}
Our values of $C_{\rm S}$, 0.171 and 0.248, correspond to 
the opening angle of the Raman-scattering region 
$\theta_{\rm R} = 48.9^{\circ}$ and $59.7^{\circ}$ for the 
spectra from 2006 April and 2007 July. For a given STB model, 
the parameter $X_{\rm H^{+}}$ is related to the 
angle between the binary axis and the asymptote to 
the ionization boundary, $\theta_{\rm a}$ 
(see Figure~\ref{fig:tha}), as 
\begin{equation}
   X_{\rm H^{+}} = \lim_{u\to \infty}f(u,\theta) = f(\theta_{a}). 
\label{eq:tha}
\end{equation}

The covering factor of the total \hi\ zone is larger than 
$C_{\rm S}$, i.e. $\theta_{\rm a} > \theta_{\rm R}$, because 
the scattering region represents only a part of the whole \hi\ 
region, which is optically thick for Raman scattering. 
In other words, the Raman-scattering \hi\ region is located 
inside the whole \hi\ zone because its angular extension 
$\theta_{\rm R}$ is bounded by the column densities 
\begin{equation}
  N_{\rm H I}(\theta) > 1.24\times 10^{21}\rm cm^{-2} 
\label{eq:nthR}
\end{equation}
(see Section~3.2.2). 
Therefore, to find the ionization boundary of the total \hi\ 
zone (i.e. the $X_{\rm H^{+}}$ parameter in the STB model) 
from the size of the Raman-scattering \hi\ zone, we have 
to determine $X_{\rm H^{+}}$ as a function of $\theta_{\rm R}$ 
taking into account the condition (\ref{eq:nthR}). 
For this purpose we reconstruct the relation, 
\begin{equation}
   X_{\rm H^{+}} = \lim_{u\to u_{\rm R}}f(u,\theta) 
                 = f(\theta_{\rm R}), 
\label{eq:thr}
\end{equation}
where $u_{\rm R}$ is the distance between the WD and the 
\hi/\hii\ interface in the line $\theta_{\rm R}$, which satisfies 
the condition, 
\begin{equation}
  N_{\rm H I}(\theta_{\rm R}) = 
              \int^{\infty}_{u_{\rm R}}n_{\rm H}(u,\theta_{\rm R})\,
             {\rm d}u = 1.24\times 10^{21}\rm cm^{-2} .
\label{eq:Nhi}  
\end{equation}
The concentrations $n_{\rm H}(u,\theta)$ are calculated according 
to Equations~(\ref{eq:cont})--(\ref{eq:vr}) along the line $\theta$ 
throughout the \hi\ region to infinity (in practice to $u = 1000$). 
In the calculation, $\dot{M}$ in Equation~(\ref{eq:cont}) is 
parameterized by $X_{\rm H^{+}}$ (Equation~(\ref{eq:X})) as 
\begin{equation}
   {\dot M} = \left (\frac{\zeta}{X_{\rm H^{+}}} \right )^{1/2}.
\label{eq:xdotm}
\end{equation}
For V1016~Cyg, $p = 1.15\times 10^{14}$\,cm, as given by the 
orbital period of 15.1 years \citep{pea02} and the total mass 
of the binary, 
$\equiv 2$\mo, 
$L_{\rm H} = 1.73\times10^{48}~\rm s^{-1}$, 
$\alpha_{\rm B}({\rm H},T_{\rm e}) = 
                        1.43\times 10^{-13}$\cmtt$\rm s^{-1}$ 
(for $T_{\rm e} = 20,000$\,K), 
$a({\rm He}) \equiv 0.15$ \citep{nuv87,mn94}, and 
$v_{\infty} = 11$\kms\ \citep{lk07}, 
yield $\zeta = 1.05\times 10^{41}$ (in units of cm\,g\,s). 
In the $\beta$-law wind (Equation~(\ref{eq:vr})) we used the average 
radius of the Mira-variable, $R_{\rm g} = 407\pm 73$\ro, as the results from 
the P-L relation of Mira variables in the LMC \citep[][]{w+09,w+13}, 
the period of 474 days \citep[][]{pea02}, and its average effective 
temperature of $2700\pm 200$\,K (from our model SED; 
in preparation). Furthermore, we adopted $\beta = 2.5$ \citep{sch85} 
and $v_{0} = 4.0\pm 0.2$\kms\ ($\approx$ the sound speed in 
the giant's atmosphere), where uncertainties in $R_{\rm g}$ 
and $v_{0}$ reflect those found in the temperature of the cool component. 

Figure~\ref{fig:xthr} shows the function $X_{\rm H^{+}}(\theta_{R})$ 
calculated according to Equation~(\ref{eq:thr}). 
Our values of $\theta_{\rm R} = 48.9^{\circ}$ and $59.7^{\circ}$, 
as determined from the spectra in 2006 April and 2007 July, 
correspond to $X_{\rm H^{+}} = 6.6$ and 3.5 for the $\beta$-law 
wind and 3.3 and 2.0 for the constant wind, respectively. 
Corresponding ionization boundaries for the $\beta$-law wind 
are shown in Figure~\ref{fig:tha}. 
Finally, accordig to Equation~(\ref{eq:xdotm}) we obtained 
$\dot{M} = 2.0$ and $2.7\times 10^{-6}$\myr\ for the $\beta$-law 
wind, and $\dot{M} = 2.8$ and $3.6\times10^{-6}$\myr\ for 
a constant velocity wind with $v_{\infty} = 11$\kms. 
The results are summarized in Table~2. 
%
%
%
%
%
%
\begin{table} 
\caption{Mass-loss Rates from the Mira-variable in V1016~Cyg as 
         determined from \heii\,1025\,\AA\ Raman 
         Scattering
        }
\label{vysl}
\begin{center}
\begin{tabular}{cccccc}
\tableline\tableline
Date                  &
$\eta$                &
C$_{\rm S}$           &
$\theta_{\rm R}$      &
$X_{\rm H^{+}}$       &
$\dot{M}/10^{-6}$    \\
                      &
$(\%)$                &
                      &
$(^{\circ})$          &
                      &
$($\mo$\,{\rm yr}^{-1})$ \\
\hline
\\
2007 Jul 31 &14.8 &0.24 &59.7 &3.5 &2.7$^{+0.2}_{-0.1}$~$^{a}$ \\
              &     &     &     &2.0 &3.6~$^{b}$ \\
2006 Apr 10 &10.2 &0.17 &48.9 &6.6 &2.0$^{+0.1}_{-0.2}$~ $^{a}$ \\
              &     &     &     &3.3 &2.8~$^{b}$ \\
2002 May 20  &17.0 &0.28 &64.0 &2.8 &3.0$^{+0.2}_{-0.2}$~ $^{c}$ \\
\hline
\end{tabular}
\end{center}
{\bf Notes.} Parameters $\eta$, C$_{\rm S}$, 
         $\theta_{\rm R}$, $X_{\rm H^{+}}$ and $\dot{M}$ 
         are given by Equations~(\ref{eq:eta2}), (\ref{eq:cs}), 
         (\ref{eq:omega}), (\ref{eq:thr}) and (\ref{eq:xdotm}), 
         respectively.\\
         $^{a}$ $\beta$-law wind.\\
         $^{b}$ Constant velocity wind of 11\kms.\\
         $^{c}$ For $\eta = 17\% $ of \cite{lee03}.	
\end{table}
\\

\section{Discussion}
\label{sect:disc}

\subsection{On the Difference Between the Angle 
            $\theta_{\rm a}$ and $\theta_{\rm R}$}
\label{sect:domega}

The angle $\Delta\theta = \theta_{\rm a} - \theta_{\rm R} > 0$ 
cut out a fraction of the \hi\ zone, where the Raman 
\heii\,$\lambda 1025 \rightarrow \lambda 6545$ conversion can 
be neglected (i.e. $\tau_{\rm Ram} < 1$). 
In our models, $\Delta\theta \la 0.2^{\circ}$, which means that 
both the Raman-scattering \hi\ region and the total \hi\ region 
are nearly identical. As a result, the function 
$X_{\rm H^{+}}(\theta_{\rm a})$ would also be nearly identical 
with the function $X_{\rm H^{+}}(\theta_{\rm R})$ plotted 
in Figure~\ref{fig:xthr}. In such a case, we can set 
$\theta_{\rm R} = \theta_{\rm a}$, and obtain $\dot{M}$ 
more easily by determining the parameter $X_{\rm H^{+}}$ 
directly with the aid of Equation~(\ref{eq:tha}). 

\begin{figure}
\begin{center}
\resizebox{\hsize}{!}
          {\includegraphics[angle=-90]{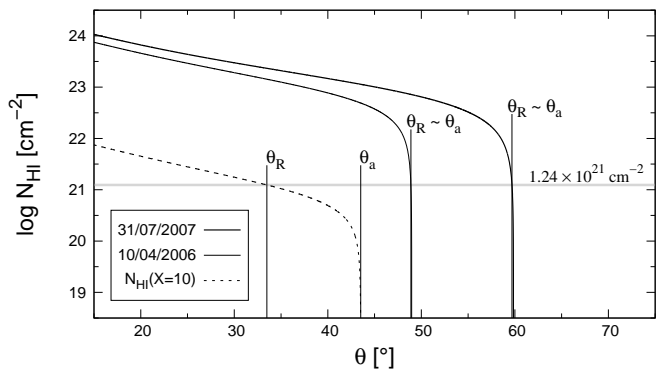}}
\end{center}                               
\caption[]{
Column densities $N_{\rm H I}(\theta)$ throughout the 
total \hi\ region above the giant photosphere 
(i.e. $\theta >13.8^{\circ}$). 
The asymptotic angle $\theta_{\rm a}$ corresponds to 
the tangent of the function $N_{\rm H I}(\theta)$, while 
the angle $\theta_{\rm R}$ denotes the intersection 
of $N_{\rm H I}(\theta)$ with the critical column density 
for \heii\,1025\,\AA\ Raman scattering 
(horizontal gray line, Equation (\ref{eq:Nhi})). 
           }
\label{fig:nhth}
\end{figure}

\begin{figure}
\begin{center}
\resizebox{\hsize}{!}
          {\includegraphics[angle=-90]{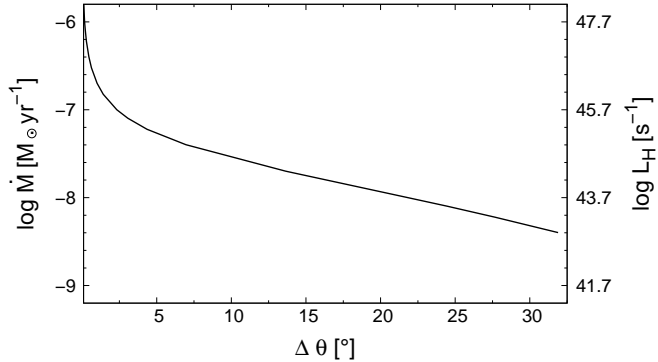}}
\end{center}                               
\caption[]{
  $\dot{M}$ and $L_{\rm H}$ that correspond to 
  $X_{\rm H^{+}} = 6.6$ (Equation~(\ref{eq:X})) as a function 
  of $\Delta \theta$ (see Section~\ref{sect:domega}).
           }
\label{fig:dth}
\end{figure}

The case $\theta_{\rm a} \gg \theta_{\rm R}$ can happen for 
low values of $\dot{M}$ and $L_{\rm H}$ that result in low 
values of column densities $N_{\rm HI}(\theta)$. 
Figure~\ref{fig:nhth} shows such an example 
(model "N$_{\rm HI}$(X=10)") together with the column densities 
of our solutions. $N_{\rm HI}(\theta)$ functions were calculated 
according to Equation~(\ref{eq:Nhi}), omitting the condition for 
the Raman-scattering limit, and for $\theta >13.8^{\circ}$, i.e., 
above the giant's photosphere. In all cases, the function 
$N_{\rm HI}(\theta)$ has a maximum just above the giant, then 
gradually decreases, and steeply drops to zero for 
$\theta \rightarrow \theta_{\rm a}$. 
The example model `N$_{\rm HI}$(X=10)' corresponds to 
$X_{H^{+}} = 10.0$ (i.e. $\theta_{\rm a} = 43.5^{\circ}$), 
$\dot{M} = 2.0\times 10^{-8}$\myr, 
$L_{\rm H} = 2.7\times 10^{44}$\,s$^{-1}$, and other 
parameters as above. 
Its $N_{\rm HI}(\theta)$ function crosses the limiting 
value at $\theta_{\rm R} = 33.5^{\circ}$, i.e., 
$\Delta\theta = 10.0^{\circ}$.
Figure~\ref{fig:dth} demonstrates a more general 
case, where the angle $\Delta\theta$ is a function of 
$\dot{M}$ and $L_{\rm H}$, which give the same value of 
$X_{\rm H^{+}} = 6.6$. The figure shows that massive winds 
flowing at rates of $\sim 10^{-6}$\myr\ have very small 
$\Delta\theta$ for \heii\,1025\,\AA\ scattering. 
%
%
%
%
%
%

\subsection{Comparison with Previous $\dot{M}$}

Similar values of the mass-loss rate from the giant in 
V1016~Cyg were derived by different methods. 
Based on the far-IR colors, as measured during the IRAS survey, 
\cite{kfs88} derived $\dot{M} = 6.3\times10^{-6}$\myr\ for 
the distance of 2.1\kpc. 
Measuring the flux density at 3.6\,cm, 
\cite{skt93} determined $\dot{M} = 1.3\times10^{-5}$\myr\ 
for the distance of 3.9\kpc. 
In both cases a constant wind velocity of 30\kms\ was assumed. 
The latter value was obtained within the context of the STB 
model, where $\dot{M} \propto v_{\infty}$. This allows us to 
convert $\dot{M}$ to $4.8\times10^{-6}$\myr\ for 
$v_{\infty} = 11$\kms. 
We note that $\dot{M}$ values derived from the radio emission 
can be overestimated because of the presence of the ionized 
wind from the hot component \citep[e.g.][]{sk06}. 
By modeling the infrared excess emission of Mira variables 
in globular clusters, \cite{frogel+88} derived typical 
values of $\dot{M}$ to be of $5\times10^{-6}$\myr. 

In our approach we used the binary separation $p = 7.7$\,AU 
(Section~\ref{sect:x}). 
Analyzing \textit{Hubble Space Telescope} images of
V1016~Cyg, \cite{bak02} estimated 
a projected binary separation of $84\pm 2$\,AU (a distance 
of 2\kpc\ was assumed), which corresponds to an extremely 
long orbital period of $\sim$544\,years. 
For a comparison, recalculating $\dot{M}$ for $p = 84$\,AU 
by our method (keeping the covering factor unchanged) yields 
$\dot{M} = 9.1\times10^{-6}$ and $1.2\times10^{-5}$\myr\ for 
the spectrum from 2006 and 2007, respectively. 
Thus, a significant enlargement of $p$ by a factor of $\sim$11 
increases $\dot{M}$ with a factor of $\sim$4.5 only. 

Finally, \cite{lee03} determined the efficiency of the 
\heii\,$\lambda 1025 \rightarrow \lambda 6545$ conversion 
to $\eta = 0.17$ (i.e., $C_{\rm S} = 0.28$ and 
$\theta_{\rm R} = 64^{\circ}$) from the spectrum obtained 
on 2002 May 20. 
Applying our method to this value, we obtained 
$X_{\rm H^{+}} = 2.8$ and $\dot{M} = 3.0\times10^{-6}$\myr, 
which is comparable to our results (see Table~2). 
\subsection{A Velocity Dispersion of the Giant's Wind}
Figure~\ref{fig:isol} shows a comparison of the Raman-scattered 
\heii\,6545\,\AA\ line with the ordinary \heii\,6527\,\AA\ line. 
The weakness of the $\lambda$6545 emission does not 
allow us to distinguish any more complex structures of its profile. 
Therefore, we matched it by a single Gaussian curve. 
Although there is no appropriate original \heii\,1025\,\AA\ 
line,\footnote{The \heii\,1025\,\AA\ line is severely blended 
with the geocoronal Ly$\beta$ emission on the only \textsl{FUSE} 
spectrum, No. A1340302000} 
we can still compare the $\lambda$6545 line profile with other 
\heii\ lines because they are formed in the same He$^{++}$ 
region around the WD and thus are similar in the profile.

We determined the FWHM of the Raman feature to 14.5 and 
15.2\,\AA\ (Table~1), which, according to Equation~(\ref{eq:dl}), 
converts to the velocity widths $\Delta v\sim$104 and 
$\sim$109\kms. For \heii\,6560\,\AA, $\Delta v\sim 91$ and 
$\sim 82$\kms\ in 2006 and 2007. The velocity width of the 
\heii\,6527\,\AA\ line is $\sim 87$\kms\ in both spectra. 
Hence, $\Delta v$ of the Raman emission is larger by 
$\approx 20$\kms. Assuming a single-peaked profile of the 
original \heii\,1025\,\AA\ line, this additional broadening 
can be attributed to the Doppler effect produced by the motion 
of the scattering H$^0$-atoms relative to the He$^{++}$ region. 
Basically, the neutral part of the wind between the two stars 
moves toward the He$^{++}$ region and thus produces blue shifted 
Raman photons, while the outer wind region moves away from it, 
producing Raman photons in the red line wing \citep[][]{sea99}. 
This result suggests a dispersion velocity of the giant's wind 
to be of $\approx 20$\kms, which can be interpreted as double that
of $v_{\infty}$ because it is consistent with a typical 
terminal velocity of the wind from Mira variables 
\citep[e.g.][]{schild89,vw93}.

However, future repeated observations that are also made for other 
symbiotic Miras and for other Raman-scattered \heii\ lines 
should allow us to obtain more accurate results.

\section{Conclusions}
\label{sect:con}
In symbiotic binaries the \heii\ emission is located predominately 
near the hot component. A part of its line photons can be 
converted by atomic hydrogen in the neutral part of the giant's 
wind into Raman photons. Within the STB model (Section~3.3.1) 
the neutral region has a conical shape, whose opening angle is 
given by the flux of hydrogen-ionizing photons and the mass-loss 
rate from the giant. Therefore, by knowing the fundamental 
parameters of the hot component, the mass-loss rate can be 
probed via Raman scattering. 
The key parameter is the efficiency $\eta$ of this process, 
because it is a function of the opening angle of the neutral 
zone. The method is described in Section~3.3. 

In this paper we investigated Raman 
\heii\,$\lambda1025\rightarrow\lambda6545$ conversion 
in the spectrum of D-type symbiotic star V1016~Cyg. 
To isolate the Raman 6545\,\AA\ emission and to derive 
$\eta$ of the corresponding \heii\,1025\,\AA\ scattering, 
we proceeded in a way that was similar \cite{lee03}. 
In our spectra from 2006 April and 2007 July, we derived 
$\eta = $0.102 and 0.148. 
Using our method, we determined the corresponding 
$\dot{M} = 2.0^{+0.1}_{-0.2}\times10^{-6}$ and 
$2.7^{+0.2}_{-0.1}\times10^{-6}$\myr. The parameters are 
summarized in Table~2. 
Our quantities of 
$\dot{M}$ are well comparable with those obtained by 
other methods (see Section~4.2). 

Deriving $\dot{M}$ by using \heii\,1025\,\AA\ scattering 
has two main advantages: 
(i) 
a large $\sigma_{\rm Ram}$ (i.e., a low Raman-limiting 
\hi\ column density) allows us to simplify the method by setting 
$\theta_{\rm R} = \theta_{\rm a}$ for sufficiently massive 
winds (Section~\ref{sect:domega}, Figure~\ref{fig:nhth}); and
(ii) 
the efficiency of \heii\,1025\,\AA\ scattering can be 
derived just from the optical spectrum (Section~\ref{sect:eff}). 

The relative simplicity and accuracy of our method 
can be used to test, for example, a possible dependence of 
$\dot{M}$ on the pulsation phase in symbiotic Miras. 
An extension of the method to other Mira--WD binary systems
may probe theoretical predictions of $\dot{M}$ at different
stages of AGB evolution. 

\acknowledgments
We thank the anonymous referee for useful comments and 
Theodor Pribulla for the acquisition of the optical spectra that 
we used in this contribution. 
This article was created through the realization of the project ITMS
No.~26220120009, based on the supporting operational Research   
and development program financed by the European Regional
Development Fund, which is also supported by the Slovak Academy 
of Sciences under a grant VEGA No. 2/0002/13. 

\textit{Facility}: \facility{David Dunlap Observatory}.

\end{document}